\documentclass{article}
\usepackage{dcase2026_techrep,amsmath,graphicx,url,times,booktabs, tabularx}
\usepackage{algorithm}
\usepackage{algpseudocode}
\usepackage{amssymb}

\title{Adaptive Diversity-Uncertainty Active Learning with Redundancy Control for Bioacoustic Event Classification}

\name{Gabriel Dubus$^{1*}$,
       Hugo Magaldi$^{1}$,
       Anatole Gros-Martial$^{2}$, 
       }

 \address{$^1$ Eco-Anthropologie, Mus\'eum National d'Histoire Naturelle, UMR7206, CNRS, Paris, France,\\
         $^2$ Centre d'Etudes Biologiques de Chizé, UMR 7372, CNRS, La Rochelle Université, France, \\
         $^*$ Corresponding author: Gabriel Dubus: gabriel.dubus@mnhn.fr \\ 
  }

\begin{document}
\ninept
\maketitle
\begin{abstract}
Active learning is a promising framework for reducing annotation costs in large-scale bioacoustic monitoring, where expert labeling is expensive and data distributions are highly heterogeneous across environments. However, existing sample selection strategies often rely on static criteria that do not adapt to the evolving reliability of model predictions during training. This limitation can lead to suboptimal exploration–exploitation trade-offs and redundant sample selection.\\
We propose an active learning strategy for multilabel bioacoustic event classification that jointly models predictive uncertainty, embedding-space diversity, and intra-batch redundancy. The method introduces an adaptive weighting scheme that progressively shifts from diversity-driven exploration in high-uncertainty regimes toward uncertainty-driven exploitation as the model becomes more confident, reflecting the increasing reliability of the classifier. To further improve annotation efficiency, a greedy Maximum Marginal Relevance (MMR) procedure is used to enforce diversity among selected samples within each acquisition batch.\\
We evaluate the proposed approach within the BioDCASE 2026 Task 4 active learning framework on terrestrial (BirdSet) and marine (ATBFL) benchmarks using pretrained audio embeddings and a fixed annotation budget. Experimental results show consistent improvements in learning efficiency and competitive in terms of macro mean Average Precision (mAP) and Area Under the Learning Curve (AULC) across heterogeneous acoustic domains. The gains are particularly pronounced on structured terrestrial soundscapes, while performance remains competitive under noisier marine conditions.\\
These findings demonstrate that adaptive acquisition strategies combining uncertainty estimation, embedding-space diversity, and redundancy-aware batch construction provide an effective and robust solution for annotation-efficient bioacoustic learning.
\end{abstract}

\begin{keywords}
Active Learning,
Bioacoustics,
Multilabel Classification,
Sample Selection,
Uncertainty Sampling,
Diversity Sampling
\end{keywords}
\section{Introduction}

Large-scale bioacoustic monitoring relies on passive acoustic systems that continuously generate massive volumes of unlabeled audio data in both terrestrial and marine environments. Although these recordings contain rich ecological information, only a small fraction can be manually annotated due to the high cost and expertise required for reliable labeling \cite{stowell_computational_2022}. These annotations are typically required to train machine learning models for acoustic event detection and classification.\\
This imbalance between data availability and annotation budget raises a central question:
\textit{given a large pool of unlabeled acoustic data, how should samples be selected for annotation in order to maximize model performance under a strict labeling budget?}\\
This work was submitted to the BioDCASE 2026 Task 4 workshop challenge on Active Learning for Bioacoustics, which specifically addresses annotation-efficient learning across terrestrial and marine acoustic domains. The challenge provides a standardized active learning framework, pretrained PerchV2 embeddings, and fixed classifier training procedures, requiring participants to focus exclusively on the design of acquisition strategies under a constrained annotation budget.\\
Active Learning (AL) provides a principled framework to address this challenge by iteratively selecting the most informative samples for labeling \cite{ren_survey_2022}. AL methods aim to optimize acquisition strategies in order to improve model performance while minimizing annotation effort \cite{tharwat_survey_2023}. However, designing effective acquisition functions remains challenging, particularly in bioacoustic settings where signals are multilabel, highly imbalanced, and strongly domain-dependent across recording locations \cite{stowell_computational_2022, mcewen_active_2024}.\\
Within the BioDCASE Task 4 setting, methods are evaluated on both BirdSet \cite{rauch_birdset_2025}, a large-scale terrestrial avian benchmark, and ATBFL (Acoustic Trends Blue Fin Library) \cite{miller_open_2021}, a marine bioacoustic dataset derived from passive acoustic monitoring datasets in the Southern Ocean. These datasets exhibit substantially different acoustic structures, label densities, and noise characteristics, making cross-domain generalization a central challenge for AL methods.\\
In this work, we study pool-based AL under a fixed annotation budget using pretrained acoustic embeddings. We propose a unified acquisition strategy that jointly models predictive uncertainty, embedding-space diversity, and batch-level redundancy control. The method introduces an adaptive weighting scheme driven by global model confidence over the unlabeled pool, progressively shifting from diversity-driven exploration in high-uncertainty regimes toward uncertainty-driven exploitation as the model becomes more confident. A greedy Maximum Marginal Relevance (MMR) procedure is used to ensure diversity within each acquisition batch.\\
\textbf{Contributions:}
\begin{itemize}
    \item We propose a unified AL strategy combining predictive uncertainty, embedding-space diversity, and batch-level redundancy reduction with a global-confidence adaptive weighting scheme that dynamically balances exploration and exploitation during training.
    \item We integrate a greedy Maximum Marginal Relevance (MMR) mechanism to reduce redundancy within acquisition batches.
    \item We evaluate the method on BirdSet and ATBFL under a fixed annotation budget, demonstrating consistent improvements across heterogeneous bioacoustic domains.
\end{itemize}

\section{Materials and Datasets}

We evaluate our method on the BioDCASE 2026 Task 4 benchmark, which includes both terrestrial and marine bioacoustic datasets. The task is formulated as a pool-based active learning setting, where only the sample acquisition strategy is learned while the classifier and training pipeline remain fixed.\\

\subsection{Active Learning Framework}

Evaluation is conducted using the BaseAL framework. At each active learning cycle, a batch of samples is selected from an unlabeled pool, their labels are automatically revealed (\textit{oracle labeling}), and added to the labeled set. The classifier is then retrained on the updated labeled set. This process is repeated until a fixed annotation budget of 500 samples is reached.\\
All methods operate on precomputed 5-second audio embeddings extracted using the PerchV2 model. These embeddings serve as fixed representations for training a randomly initialized multilabel classification head, ensuring a controlled comparison of sampling strategies.

\subsection{Datasets}

The benchmark consists of two complementary bioacoustic datasets designed to evaluate generalization across terrestrial and marine acoustic environments. We use precomputed embeddings extracted using the PerchV2 model \cite{rauch_biodcase_2026, kurinchi-vendhan_biodcase_2026}.

\subsubsection{BirdSet (Terrestrial)}

BirdSet is a large-scale avian bioacoustic dataset containing recordings from diverse geographic regions \cite{rauch_birdset_2025}. For this task, we use three of the eight validation sets of Birdset: HSN, POW, and UHH. These subsets vary significantly in class cardinality (19,41 and 25 respectively), and label density (average labels per sample are 0.524, 2.833 and 1.058 respectively), making them well suited for evaluating robustness across structured soundscapes.

\subsubsection{ATBFL (Marine)}

ATBFL is a marine bioacoustic dataset derived from long-term monitoring of Antarctic whale populations \cite{miller_open_2021}. It consists of multiple site-year subsets representing distinct acoustic environments.\\
Compared to BirdSet, ATBFL exhibits higher noise levels, making it a more challenging domain for embedding-based similarity and diversity modeling.

\subsection{Evaluation Protocol}

All methods are evaluated under a fixed annotation budget of 500 samples per subset. Performance is measured using the Area Under the Learning Curve (AULC) for macro mAP, averaged over 5 independent runs. 

\section{Method}

We propose ADU-MMR (Adaptive Diversity–Uncertainty + Maximum Marginal Relevance), an active learning strategy that combines predictive uncertainty, embedding-space diversity, and batch-level redundancy reduction. The method operates on fixed audio embeddings and follows an adaptive exploration–exploitation schedule throughout the annotation process.

\subsection{Problem Setting}

Let $L_t$ and $U_t$ denote the labeled and unlabeled sets at iteration $t$. At each active learning step, a batch $S_t \subset U_t$ of size $B$ is selected for annotation. Each sample $x_i$ is represented by a fixed pretrained embedding extracted using PerchV2. The classifier produces a multilabel probability vector $p_i = (p_{i,1}, \ldots, p_{i,C})$, where $C$ is the number of classes. After annotation, selected samples are added to the labeled set and the model is updated.

\subsection{Uncertainty Estimation}

Predictive uncertainty is estimated using the sum of binary entropies computed from the multilabel class probabilities:

\begin{equation}
h_i = -\sum_{c=1}^{C}\left[p_{i,c}\log(p_{i,c}+\epsilon) + (1-p_{i,c})\log(1-p_{i,c}+\epsilon)\right]
\end{equation}

where $\epsilon$ is a small constant introduced for numerical stability.\\
To ensure compatibility with the diversity score, $\tilde{h}_i = \mathrm{Norm}(h_i)$ is defined as uncertainty values linearly normalized to the interval $[0,1]$ through a min-max normalization over the unlabeled pool at iteration $t$.

\subsection{Embedding-Based Diversity}

To encourage exploration of underrepresented regions of the embedding space, we compute the distance between each unlabeled sample and its nearest labeled neighbor. For each sample $x_i \in U_t$:
\begin{equation}
d_i = \min_{x_j \in L_t} ||x_i - x_j||_2^2.
\end{equation}
The resulting distances are normalized to obtain $\tilde{d}_i$. When no labeled samples are available, all samples are assigned a diversity score of 1.

\subsection{Adaptive Exploration}

The relevance score of each sample is defined as:
\begin{equation}
s_i = \alpha_t \tilde{h}_i + (1 - \alpha_t)\tilde{d}_i.
\end{equation}

The adaptive coefficient $\alpha_t$ is driven by the global uncertainty of the model over the unlabeled pool. We compute:
\begin{equation}
H_t = \frac{1}{|U_t|} \sum_{x_i \in U_t} \tilde{h}_i.
\end{equation}

The coefficient is then defined as:
\begin{equation}
\alpha_t = 0.5 \cdot \left(\frac{\max(0, \tau - H_t)}{\tau}\right)^2,
\end{equation}

This formulation reduces reliance on uncertainty in early stages when predictions are unreliable, and gradually increases its influence as the model becomes more confident. The threshold $\tau$ has been set empirically at $\tau=0.05$ to ensure that uncertainty is only exploited once the model reaches a sufficiently low-entropy regime, where predictions become more reliable and uncertainty estimates more informative.

\subsection{Batch Selection via Maximum Marginal Relevance}

Selecting the top-$B$ samples according to $s_i$ can lead to redundant acquisitions. To mitigate this issue, we employ a greedy Maximum Marginal Relevance (MMR) procedure. This formulation follows the classical Maximum Marginal Relevance framework of Carbonell and Goldstein \cite{carbonell_use_1999}. Embeddings are $\ell_2$-normalized and pairwise cosine similarities are computed. At each step, the candidate maximizing:

\begin{equation}
\mathrm{MMR}(x_i) = s_i - \lambda \max_{x_j \in S_t} \cos(x_i, x_j)
\end{equation}

where $S_t$ denotes the set of samples selected so far within the current greedy construction of the acquisition batch at iteration $t$. The procedure is initialized with $S_t = \emptyset$ and proceeds by sequentially adding the sample maximizing the MMR criterion until $B$ samples have been selected. \\
The MMR trade-off parameter was set to $\lambda=0.3$. This value was selected empirically based on preliminary experiments, providing a good balance between score $s_i$ maximization and redundancy reduction across the considered datasets.\\
The first selected sample corresponds to the highest acquisition score. Subsequent selections are penalized according to their similarity with previously selected samples, promoting diversity within the acquisition batch.

\begin{algorithm}[h]
\caption{Adaptive Diversity--Uncertainty Active Learning}
\label{alg:al}
\begin{algorithmic}[1]
\For{each iteration $t$}
\State Compute normalized uncertainty scores $\tilde{h}_i$
\State Compute normalized diversity scores $\tilde{d}_i$
\State Compute $\alpha_t$
\State Compute acquisition scores
$s_i=\alpha_t \tilde{h}_i + (1-\alpha_t)\tilde{d}_i$
\State Initialize $S_t \leftarrow \emptyset$
\While{$|S_t| < B$}
\State Compute MMR scores
$\mathrm{MMR}(x_i)=s_i-\lambda \max_{x_j \in S_t} \mathrm{cos}(x_i,x_j)$
\State Select the highest-scoring sample
\State Add it to $S_t$
\EndWhile
\State Query annotations and update $L_t$
\EndFor
\end{algorithmic}
\end{algorithm}

\section{Experiments}

We evaluate ADU-MMR on the BirdSet (terrestrial) and ATBFL (marine) bioacoustic datasets using pretrained PerchV2 embeddings. Both datasets exhibit strong class imbalance and multilabel annotations, making them suitable benchmarks for active learning under realistic acoustic conditions.\\
The annotation budget is fixed to 500 samples with a batch size of 25, resulting in 20 active learning iterations. All experiments are repeated over 5 independent runs, and results are reported as mean $\pm$ standard deviation.\\
We compare the proposed method against the following baselines: Random sampling, Margin-based uncertainty sampling, CoreSet selection \cite{sener_active_2018} and TypiClust (diversity-aware baseline).\\

\subsection{Results}
Table \ref{tab:main_results} reports AULC and mAP scores on BirdSet and ATBFL for ADU-MMR and baselines. \\
Averaged across datasets, ADU-MMR achieves the best performance with 0.594 mAP and 0.505 AULC, outperforming all competing methods. These results indicate that the proposed strategy provides a consistent trade-off between exploration and exploitation across heterogeneous acoustic domains.\\
\begin{table}[t]
\centering
\small
\setlength{\tabcolsep}{4pt}
\begin{tabular}{lcc}
\toprule
\textbf{Method} & \textbf{AULC (mAP macro)} & \textbf{Best mAP} \\
\midrule
ADU-MMR & \textbf{0.505} & \textbf{0.590} \\
CoreSet & 0.479 & 0.583 \\
Margin Sampling & 0.422 & 0.532 \\
Random & 0.405 & 0.482 \\
\bottomrule
\end{tabular}
\caption{Overall performance averaged across BirdSet and ATBFL datasets over 5 runs per dataset, then averaged across datasets. The proposed method outperforms baselines in both AULC and mAP.}
\label{tab:main_results}
\end{table}
Table \ref{tab:per_dataset_results} reports per-dataset results of ADU-MMR and baselines. On BirdSet, ADU-MMR consistently improves performance across all subsets. On HSN, it significantly outperforms CoreSet (0.632 vs 0.556 AULC), indicating strong gains in both uncertainty- and diversity-driven regimes. Similar improvements are observed on POW, where it achieves 0.482 AULC, outperforming both CoreSet and random sampling. On UHH, gains are smaller but consistent, with ADU-MMR achieving the best AULC and mAP. On ATBFL, all methods show comparable performance, Random achieving best scores.

\begin{table}[t]
\centering
\small
\resizebox{\linewidth}{!}{
\begin{tabular}{lcccc}
\toprule
\textbf{Dataset} & \textbf{ADU-MMR} & \textbf{CoreSet} & \textbf{Margin} & \textbf{Random} \\
\midrule

\multicolumn{5}{c}{\textbf{AULC (mAP macro)}} \\
\midrule

ATBFL & 0.454 & 0.462 & 0.441 & \textbf{0.464} \\
HSN & \textbf{0.632} & 0.556 & 0.444 & 0.382 \\
POW & \textbf{0.482} & 0.476 & 0.461 & 0.438 \\
UHH & \textbf{0.443} & 0.421 & 0.341 & 0.337 \\

\midrule
\multicolumn{5}{c}{\textbf{Best mAP}} \\
\midrule

ATBFL & 0.484 & 0.489 & 0.483 & \textbf{0.494} \\
HSN & \textbf{0.716} & 0.696 & 0.600 & 0.477 \\
POW & \textbf{0.617} & 0.608 & 0.607 & 0.559 \\
UHH & \textbf{0.543} & 0.541 & 0.440 & 0.399 \\

\bottomrule
\end{tabular}
}
\caption{Per-dataset performance of ADU-MMR and baselines in terms of AULC (macro mAP) and best achieved mAP. Best values per row are highlighted in bold.}
\label{tab:per_dataset_results}
\end{table}

% ------------------------------------------------------------
\subsection{Effect of the adaptive weighting scheme}

\begin{figure}
    \centering
    \includegraphics[width=0.99\linewidth]{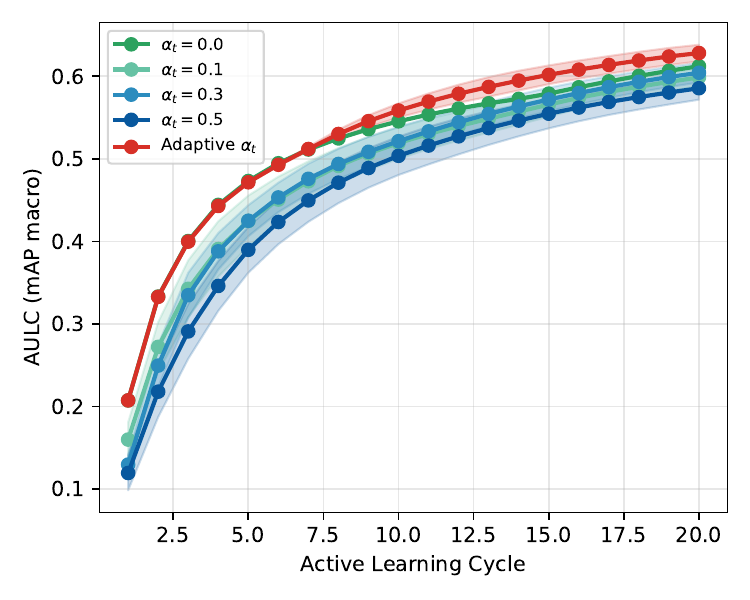}
    \caption{Evolution of AULC across active learning cycles for different fixed values of $\alpha$ (0, 0.1, 0.3, 0.5) as well as the proposed adaptive schedule, results for BirdSet-HSN dataset. The adaptive schedule consistently outperforms all fixed-weight configurations, particularly during early active learning cycles.}
    \label{fig:alpha_eval}
\end{figure}

We analyze the impact of the adaptive weighting parameter $\alpha_t$ on the active learning performance. Figure \ref{fig:alpha_eval} reports the evolution of AULC across active learning cycles for different fixed values of $\alpha$ (0, 0.1, 0.3, 0.5) as well as the proposed adaptive schedule, computed with the example of the BirdSet-HSN dataset.

Overall, the adaptive strategy consistently outperforms all fixed weighting configurations across active learning cycles. The largest performance gap is observed in the early stages, where high values of $\alpha$ underperform due to unreliable uncertainty estimates from a limited labeled set. In this regime, diversity-based selection is more effective than uncertainty-driven sampling.\\
At the beginning of the active learning process, the adaptive strategy behaves similarly to the $\alpha = 0.0$ configuration, as it assigns a low value to $\alpha_t$ and prioritizes diversity. As training progresses, the two strategies diverge: the fixed $\alpha = 0.0$ configuration saturates, while the adaptive approach continues to improve. This suggests that uncertainty estimates become increasingly reliable as the model is exposed to more labeled data, making uncertainty-based acquisition progressively more beneficial.\\
These results highlight the importance of dynamically balancing uncertainty and diversity rather than relying on a fixed weighting scheme.\\

\section{Discussion}
The results highlight a clear dependency of active learning performance on dataset structure and embedding quality. On BirdSet, the proposed method consistently outperforms all baselines, indicating that pretrained embeddings provide a meaningful geometric structure that can be effectively exploited by both uncertainty and diversity-based acquisition strategies.\\
In contrast, results on ATBFL are more limited, with all methods exhibiting similar performance. This suggests that sample selection strategies have a reduced impact in this setting, where performance gains from additional labeled data remain marginal across iterations.\\
ATBFL is a marine bioacoustic dataset composed of low-frequency recordings of large cetaceans in the Southern Ocean. Several target signals occur at very low frequencies (e.g., blue whale Z-calls around 25 Hz and fin whale calls around 20 Hz). However, the pretrained PerchV2 embeddings may not fully capture such low-frequency acoustic content, which could limit their representational adequacy in this domain. This likely contributes to the limited performance gains observed across active learning strategies on ATBFL. In such cases, when the underlying representation space does not adequately separate relevant acoustic events, improvements from sample selection strategies may be inherently constrained.\\
We further analyze the role of the adaptive weighting parameter $\alpha_t$. Early in the active learning process, low values of $\alpha_t$ are beneficial due to unreliable uncertainty estimates from a classifier trained on limited labeled data. In this regime, diversity-based selection provides more stable exploration of the embedding space. As labeling progresses, uncertainty estimates become more reliable, allowing the method to progressively shift toward uncertainty-driven acquisition \cite{soviany_curriculum_2022}.\\

\section{Conclusion}
The proposed method combines uncertainty estimation, embedding-based diversity, and batch-level redundancy reduction within a unified active learning framework. It dynamically balances exploration and exploitation through an adaptive weighting scheme and enforces diversity via greedy MMR-based selection.\\
Experiments on BirdSet and ATBFL show consistent improvements over baselines, with particularly strong gains in structured terrestrial environments. Overall, the results demonstrate the effectiveness of adaptive acquisition strategies for multilabel bioacoustic active learning under strict annotation budgets.\\

\section{ACKNOWLEDGMENT}
\label{sec:ack}
The author thanks the BioDCASE 2026 Task 4 organizers for providing the BaseAL framework and curated datasets.

%\newpage
% -------------------------------------------------------------------------
% Either list references using the bibliography style file IEEEtran.bst
\bibliographystyle{IEEEtran}
\bibliography{refs}

\end{document}